\documentclass[twocolumn,notitlepage,pra,superscriptaddress,longbibliography]{revtex4-1}
\usepackage{textcomp}
\usepackage{times}
\usepackage{graphicx}
\usepackage{float}
\usepackage{latexsym,amsmath,amssymb,bm,euscript}
\usepackage{color}
\usepackage{subfigure}
\usepackage{epstopdf}
\usepackage[colorlinks=true,linkcolor=blue,citecolor=blue]{hyperref}
\usepackage{hyperref}
\usepackage{soul}
\usepackage[normalem]{ulem}
\usepackage{mathrsfs}
\usepackage{amsmath}
\usepackage{xspace}
\usepackage{CJKutf8}

\graphicspath{{./Figs/}}

\begin{document}
\begin{CJK}{UTF8}{<font>}
\title{High-order exceptional point in a nanofiber cavity quantum electrodynamics system}

\author{Zigeng Li}
\affiliation{School of Physics, Beihang University, Beijing 100191, China}

\author{Xiaomiao Li}
\affiliation{School of Physics, Beihang University, Beijing 100191, China}

\author{Xiaolan Zhong}
\email{zhongxl@buaa.edu.cn}
\affiliation{School of Physics, Beihang University, Beijing 100191, China}

\begin{abstract}
We present an all-fiber emitter-cavity quantum electrodynamics (QED) system which consists of two two-level emitters and a nanofiber cavity.
Our scheme makes it possible to observe the higher-order exceptional points based on the coupling between the emitters and the nanofiber cavity.
The effective gain of this cavity can be obtained by weakly driven to the nanofiber cavity via two identical laser fields, which will realize coherent perfect absorption (CPA) in the implementation of the experiments.
Under the experimental feasible parameters, the Hamiltonian of this system is in the condition of pseudo-Hermiticity, which means that its eigenvalues can be made of one real and a pair of complex conjugates, or be all real.
By controllably tuned the ratio of the two emitter-cavity coupling strengths, and the ratio of the decay rates of the emitters, we can discover both the three-order exceptional point (EP$_{3}$) and the second-order exceptional point (EP$_{2}$) without parity-time symmetry in our emitter-cavity system.
These results can also be demonstrated by the total output spectra and transmission spectra. We also find that the symmetric modes come into being when the coupling strength greater than the critical coupling strength at EP$_{3}$ points.
Our proposal will provide a new method to realize higher-order exceptional points based on the quantum network.

\end{abstract}

\date{\today} 
\maketitle

\section{INTRODUCTION}
The cavity quantum electrodynamics (QED) systems based on optical fiber have received growing interest in recent years due to its low propagation losses and easy to preparation in experiments \cite{RN30}.
Among the various fiber-optics systems, optical nanofibers (ONFs) have been used in different cavity QED systems owing to their conspicuous optical characteristics \cite{RN31,PhysRevLett.115.093603,PhysRevLett.122.253603,RN151}.
For example, the steep variation of the evanescent field around the ONFs lead to a gradient force on the atom, which can be used to trap it in the waist regime \cite{PhysRevLett.104.203603,PhysRevLett.109.033603}.
Compared with other systems, one of the important properties of the nanofiber-cavity QED systems is that the length of the ONFs will not affect the cooperativity \cite{RevModPhys.90.031002}.
Thus, the effective mode area is very small so that even the long length cavity with low finesse can realize the strong coupling with the atoms.
To date, the ONFs have been utilized as an important component in fiber cavity QED systems. In experiment, researchers have achieved the observation of dressed states of distant atoms and cavity dark mode in a coupled nanofiber cavity QED system \cite{PhysRevLett.122.253603,RN31}. Moreover, the ultra-strong photon blockade based on the nanofiber cavity QED system has been proved in theory \cite{PhysRevA.103.043724}.

One of the fundamental postulates of quantum mechanics is that the Hamiltonian $H$ is assumed to be Hermitian, that is $H={{H}^{\dagger }}$, where the superscript $\left( \dagger  \right)$ represents the Hermitian conjugation.
This equation will produce the real eigenvalues and ensure that the probability to find the particle somewhere is conserved.
Therefore, these systems will be described by the non-Hermitian Hamiltonian, that is $H\ne {{H}^{\dagger }}$.
In general case, this type of Hamiltonians can still possess the real spectrum if they have the property of being parity-time ($\mathcal{P}\mathcal{T}$) symmetry, which satisfy the following reciprocal relation \cite{RevModPhys.88.035002}: $\left[ \mathcal{P}\mathcal{T},H \right]=0$, where $\mathcal{P}$ and $\mathcal{T}$ obey:
$\mathcal{P}\psi (x)=\psi (-x), \mathcal{T}\psi (x)=\overset{\_}{\mathop{\psi }}\,(x)$.
According to the approach presented by Mostafazadeh \cite{RN152,RN153,RN154}, the non-Hermitian operators can be connected with corresponding adjoints by a transformation as: ${{H}^{\dagger }}=\eta H{{\eta }^{-1}}$, which be defined as pseudo-Hermitian operators, and both of the $H$ and $H={{H}^{\dagger }}$ have the same eigenvalues in the form of real or complex conjugate.

It is obvious that every non-Hermitian Hamiltonian with $\mathcal{P}\mathcal{T}$-symmetric is pseudo-Hermitian. Many new physical phenomena have been proposed based on pseudo-Hermiticity.
For example, the real spectra in non-Hermitian topological insulators \cite{PhysRevResearch.2.033391}, quantum phase transitions\cite{PhysRevE.80.021107,PhysRevE.80.026213,PhysRevA.90.012103}, Goldstone’s theorem\cite{PhysRevD.101.045014}, $\mathcal{P}\mathcal{T}$-symmetric quantum walks \cite{PhysRevA.96.032305} and quantum sensing \cite{PhysRevLett.124.020501}.
In the following of our analysis, the pseudo-Hermiticity we considered excludes both the Hermiticity and the $\mathcal{P}\mathcal{T}$-symmetry.
Traditionally, the quantum phase transition from $\mathcal{P}\mathcal{T}$-symmetric to the broken $\mathcal{P}\mathcal{T}$-symmetric take place near the critical points, where the two or more eigenvalues simultaneously coalesce and become degenerate \cite{RN156}.
These points are known as exceptional points (EPs). Owing to the EPs open pathways for new functionalities and performance, so far, various physical systems have been studied the concepts of EPs both theoretically and experimentally,
including cavity magnonics systems \cite{PhysRevB.99.214415,PhysRevB.99.054404,RN57,PhysRevApplied.13.014053,PhysRevLett.124.053602,PhysRevLett.125.147202},
cavity optomechanical systems \cite{PhysRevE.98.032201,PhysRevLett.114.253601,PhysRevApplied.12.024002,PhysRevA.96.033856},
non-Hermitian optical gyroscopes \cite{PhysRevA.104.033505},
two-level quantum dots (QDs) cavity QED systems \cite{PhysRevResearch.2.043075,PhysRevA.95.042115,PhysRevA.96.013802}
and directly or indirectly coupled microresonators \cite{RN157,PhysRevA.100.043810,PhysRevA.102.043527}.

In particular, high-order exceptional points can also appear in non-Hermitian systems. Obviously, three or more eigenvalues simultaneously coalesce at these points \cite{PhysRevA.101.033820,PhysRevLett.127.186601,RN159,RN160,RN161,RN162,RN163}. Although it is more complex than second-order EPs in theory, but it can display richer physical phenomena near these points. Specifically, it can be used to enhance sensing in subwavelength resonator arrays \cite{PhysRevA.101.033820}, or enhanced Topological Energy Transfer in Magnonic Planar Waveguides \cite{PhysRevApplied.15.034050}.
And sensitivity enhancement of a sensor operated at the three-order EPs in a $\mathcal{P}\mathcal{T}$-symmetric photonic laser molecule \cite{RN158}.
To best of our knowledge, we first use the method of the pseudo-Hermiticity without  $\mathcal{P}\mathcal{T}$-symmetry and achieve the three-order exceptional points (EP$_{3}$) in an all-fiber emitter-cavity quantum electrodynamics system.

In this work, our proposal is very different from previous cavity QED systems which can be achieved  $\mathcal{P}\mathcal{T}$-symmetric \cite{PhysRevResearch.2.043075,PhysRevA.96.013802}. However, it is still a challenge to realize high-order EPs in atom-cavity QED systems and few work in terms of it.
According to our previous work \cite{PhysRevA.103.043724}, We put forward a system consisting of double two-level emitters coupled with a common nanofiber cavity. In this manuscript, the emitter including different types of two-level structures, such as cold alkali metal atoms (e.g., Rb, Cs), or quantum qubits etc.
In order to meet the conditions of pseudo-Hermiticity of the Hamiltonian, the cavity with gain is needed. We can achieve it by weakly driven the nanofiber cavity to achieve coherent perfect absorption (CPA) \cite{PhysRevA.97.053825}.
Besides the second-order exceptional points (EP$_{2}$), the EP$_{3}$ can also be found in our system. To further demonstrate our observation, we measure the total output spectra and absorption spectra of the nanofiber cavity.

The remainder of this manuscript is arranged in five parts as follows. In Sec.~\ref{Sec:2}, we first introduce a nanofiber emitter-cavity QED system with a pseudo-Hermitian Hamiltonian without $\mathcal{P}\mathcal{T}$-symmetric and discuss the origin of the EP$_{3}$.
Subsequently, in Sec.~\ref{Sec:3}, we give two different cases to achieve both second-order and third-order EPs in the system, and consider the interesting phenomenon observed near the EPs. In Sec.~\ref{Sec:4}, we measure the output spectrum to prove the Correctness of EPs and we discuss the experimental implementation in Sec.~\ref{Sec:5}. Finally, we give the conclusion in Sec.~\ref{Sec:6}.

\begin{figure}[!tbp]
\includegraphics[angle=0,width=1\linewidth]{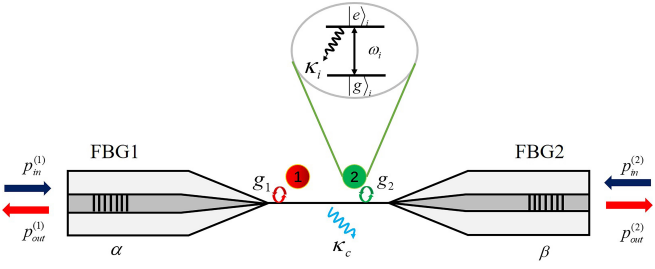}
\caption{(Color online)
Simple schematic diagram of the system, which consists of two FBG mirrors in the single-mode-fibers and a nanofiber waist. All of them form the whole Fabry-Perot nanofiber cavity system. Two two-level (ground state ${{\left| g \right\rangle }_{i}}$ and excited state ${{\left| e \right\rangle }_{i}}$) emitters with resonant frequency ${{\omega }_{i}}$ ($i=1,2$) are trapped in evanescent field of a nanofiber. The blue (red) arrows represent the input (output) field $p_{in}^{(i)}$ ($p_{out}^{(i)}$). The nanofiber cavity mode with decay rate is given by ${{\kappa }_{c}}=\alpha +\beta +{{\kappa }_{\operatorname{int}}}$, where $\alpha $ ($\beta $) and ${{\kappa }_{\operatorname{int}}}$ are the field decay rate through left (right) FBG and intracavity losses, respectively.
}
\label{Fig:1}
\end{figure}

\section{THE MODEL}
\label{Sec:2}

\subsection{The effective Hamiltonian of the system}
\label{Sec:2s}

The quantum system we consider based on a nanofiber cavity QED, which consists of two two-level emitters (with transition frequency ${{\omega }_{1}}$ and ${{\omega }_{2}}$, respectively) coupling with a Fabry Perot nanofiber cavity (with frequency ${{\omega }_{c}}$).
The nanofiber is used to form a tapered region as a nanofiber waist, which connecting the two standard optical fibers with two fiber-Bragg-grating (FBG) mirrors, thereby forming an all-fiber cavity QED. The emitters are trapped in a state-insensitive nanofiber trap
\cite{PhysRevLett.117.133603,PhysRevA.97.032509}, as illustrated in Fig.~\ref{Fig:1}.

We assume that the system is driven by the weak field, the Fabry-Perot cavity only support the single-mode (fundamental mode HE$_{11}$), this system can be well decided by the Jaynes-Cummings (JC) Hamiltonian, which can be written as:
\begin{equation}
{{H}_{sys}}={{H}_{0}}+{{H}_{\operatorname{int}}},
\label{Eq:1}
\end{equation}
where, ${{H}_{0}}/\hbar ={{\omega }_{c}}{{a}^{\dagger }}a+{{\omega }_{1}}\sigma _{1}^{\dagger }{{\sigma }_{1}}+{{\omega }_{2}}\sigma _{2}^{\dagger }{{\sigma }_{2}}$ is the total free energy of the cavity and emitters. Also, ${{a}^{\dagger }}(a)$ is the creation (annihilation) operator of the cavity field and ${{\sigma }_{i}}=\left| g \right\rangle \left\langle  e \right|$ ($\sigma _{i}^{\dagger }=\left| e \right\rangle \left\langle  g \right|,i=1,2$) is the lowing (raising) operator of the emitters.

The second term in Eq.~(\ref{Eq:1}) represents the interaction Hamiltonian between the cavity and the two-level emitters. Under the rotating-wave approximation, the ${{H}_{\operatorname{int}}}$ reads as ${{H}_{\operatorname{int}}}/\hbar =\sum\limits_{i=2}{{{g}_{i}}({{a}^{\dagger }}{{\sigma }_{i}}+a\sigma _{i}^{\dagger })}$, where ${{g}_{i}}={{d}_{i}}\sqrt{{{\omega }_{c}}/({{\varepsilon }_{0}}\hbar Al)}$ is the cavity-emitter coupling strengths.
Here, ${{d}_{i}}$ being the dipole momentum, ${{\varepsilon }_{0}}$ being the vacuum permittivity, and the Fabry-Perot cavity of length $l$ with cross-section area $A$.
The weakly probe fields $p_{in}^{(1)}$ and $p_{in}^{(2)}$ are fed into the nanofiber cavity via the left port of the system. We take the master equation to describe the dynamics of the open quantum system, which can be treated according to the Jaynes-Cummings model:
\begin{equation}
\frac{d}{dt}\rho =-i\left[ {{H}_{sys}},\rho  \right]+{{L}^{(c)}}\left[ \rho  \right]+{{L}^{(i)}}\left[ \rho  \right],
\label{Eq:2}
\end{equation}
where, ${{L}^{(i)}}\left[ \rho  \right]={{\kappa }_{i}}D[{{\sigma }_{i}}]\rho ,{{L}^{(c)}}\left[ \rho  \right]={{\kappa }_{c}}D[a]\rho$  and $D\left[ o \right]=2o\rho {{o}^{\dagger }}-{{o}^{\dagger }}o\rho -\rho {{o}^{\dagger }}o$ ($o=a,{{\sigma }_{i}}$), ${{\kappa }_{i}}$ and ${{\kappa }_{c}}$ denote the decay rates of the emitters and cavity, respectively. Therefore, we can get a series of equations:

\begin{equation}
\begin{aligned}
\label{Eq:3}
&\frac{d}{dt}a=-(i{{\omega }_{c}}+{{\kappa }_{c}})a-i{{g}_{1}}{{\sigma }_{1}}-i{{g}_{2}}{{\sigma }_{2}}+\sqrt{2\alpha }p_{in}^{(1)}+\sqrt{2\beta }p_{in}^{(2)},\\
&\frac{d}{dt}{{\sigma }_{1}}=-(i{{\omega }_{1}}+{{\kappa }_{1}}){{\sigma }_{1}}+i{{g}_{1}}(\left\langle {{\sigma }_{z}} \right\rangle a),\\
&\frac{d}{dt}{{\sigma }_{2}}=-(i{{\omega }_{2}}+{{\kappa }_{2}}){{\sigma }_{2}}+i{{g}_{2}}(\left\langle {{\sigma }_{z}} \right\rangle a),
\end{aligned}
\end{equation}

where, we take the form $o\equiv \left\langle o \right\rangle $, $p_{in}^{(1)}$ and $p_{in}^{(2)}$ represent the input fields to the system from two ports and we assume that the setup is in the weak driving limit. .
Therefore, the excited states of the emitters are not significantly populated, utilizing the mean-field approximation, that is,  $\left\langle {{\sigma }_{z}}a \right\rangle \simeq \left\langle {{\sigma }_{z}} \right\rangle a\simeq -a$.

According to the input-output theory, the relationship of the intracavity field a with the input field $p_{in}^{(i)}$ and output field $p_{out}^{(i)}$ can be written as

\begin{equation}
\label{Eq:4}
p_{in}^{(1)}+p_{out}^{(1)}=\sqrt{2\alpha }a, p_{in}^{(2)}+p_{out}^{(2)}=\sqrt{2\beta }a.
\end{equation}

When CPA occurs, there are no output fields going out through the two ports 1 and 2, i.e., $p_{out}^{(i)}=0$. In this case, Eq.~(\ref{Eq:4}) becomes

\begin{equation}
p_{in}^{(1)}=\sqrt{2{{\alpha }}}a, p_{in}^{(2)}=\sqrt{2{{\beta }}}a.
\label{Eq:5}
\end{equation}

Substituting Eq.~(\ref{Eq:5}) into Eq.~(\ref{Eq:3}), we can get

\begin{equation}
\label{Eq:6}
\begin{aligned}
&\frac{d}{dt}a=-(i{{\omega }_{c}}-{{\kappa }_{e}})a-i{{g}_{1}}{{\sigma }_{1}}-i{{g}_{2}}{{\sigma }_{2}},\\
&\frac{d}{dt}{{\sigma }_{1}}=-(i{{\omega }_{1}}+{{\kappa }_{1}}){{\sigma }_{1}}-i{{g}_{1}}a,\\
&\frac{d}{dt}{{\sigma }_{2}}=-(i{{\omega }_{2}}+{{\kappa }_{2}}){{\sigma }_{2}}-i{{g}_{2}}a,
\end{aligned}
\end{equation}

where ${{\kappa }_{e}}={{\alpha }_{1}}+{{\beta }_{2}}-{{\kappa }_{{int}}}$. Owing to the CPA of two input fields fed into the system, this cavity can be treated as an active one.

The equations in Eq.~(\ref{Eq:6}) can be expressed in matrix form as: $\overset{\centerdot }{\mathop{V}}\,=-i{{H}_{\text{eff}}}V$, where ${{V}^{T}}={{(a,{{\sigma }_{1}},{{\sigma }_{2}})}^{T}}$ denotes a column vector and ${{H}_{\text{eff}}}$ is the effective non-Hermitian Hamiltonian of the quantum system:
\begin{equation}       
\label{Eq:7}
{{H}_{\text{eff}}}=\left(                 
\begin{array}{ccc}   
{{\omega }_{c}}+i{{\kappa }_{e}} & {{g}_{1}} & {{g}_{2}}\\  
{{g}_{1}} & {{\omega }_{1}}-i{{\kappa }_{1}} & 0\\
{{g}_{2}} & 0 & {{\omega }_{2}}-i{{\kappa }_{2}}\\
\end{array}
\right),                 
\end{equation}
where ${{\kappa }_{e}}=\alpha +\beta -{{\kappa }_{\operatorname{int}}}$ denotes the effective gain of the cavity due to the CPA. In the following sections, we will discuss the pseudo-Hermiticity based on the effective Hamiltonian ${{H}_{eff}}$.

\subsection{pseudo-Hermiticity Hamiltonian}
\label{Sec:2y}
For this considered Hamiltonian, there are three eigenvalues according to mathematical calculation. Utilizing the conclusions in Ref. \cite{RN153},
${{H}_{\text{eff}}}$ becomes pseudo-Hermiticity satisfy two conditions: one case is all of the eigenvalues are real and the other case is one eigenvalue is real and others is a complex-conjugate pair. By solving $\text{Det}({{H}_{\text{eff}}}-\varpi I)=0$, i.e.,

\begin{equation}
\left|\begin{matrix}
\label{Eq:8}
   {{\omega }_{c}}+i{{\kappa }_{e}}-\varpi  & {{g}_{1}} & {{g}_{2}}  \\
   {{g}_{1}} & {{\omega }_{1}}-i{{\kappa }_{1}}-\varpi  & 0  \\
   {{g}_{2}} & 0 & {{\omega }_{2}}-i{{\kappa }_{2}}-\varpi   \\
\end{matrix}\right|=0.
\end{equation}

Here, $I$ is the identity matrix. In order to satisfy the energy-spectrum property of the pseudo-Hermitian Hamiltonian, both Eq.~(\ref{Eq:8}) and its complex-conjugate expression share the same solutions, that is $\text{Det}(H_{\text{eff}}^{*}-\varpi I)=0$:
\begin{equation}
\left|\begin{matrix}       
\label{Eq:9}
{{\omega }_{c}}-i{{\kappa }_{e}}-\varpi  & {{g}_{1}} & {{g}_{2}}\\  
{{g}_{1}} & {{\omega }_{1}}+i{{\kappa }_{1}}-\varpi  & 0\\
{{g}_{2}} & 0 & {{\omega }_{2}}+i{{\kappa }_{2}}-\varpi \\
\end{matrix}\right|=0.
\end{equation}

We compare the above two equations [Eq.~(\ref{Eq:8})) and Eq.~(\ref{Eq:9})] and extract identical parts, making the other different parts zero. Thereby the following constraints read:

\begin{equation}
\begin{aligned}
\label{Eq:10}
&({{\kappa }_{1}}+{{\kappa }_{2}})-{{\kappa }_{e}}=0,\\
&{{\kappa }_{1}}{{\delta }_{1}}+{{\kappa }_{2}}{{\delta }_{2}}=0,\\
&{{\kappa }_{e}}({{\delta }_{1}}{{\delta }_{2}}-{{\kappa }_{1}}{{\kappa }_{2}})+g_{2}^{2}{{\kappa }_{1}}+g_{1}^{2}{{\kappa }_{2}}=0,
\end{aligned}
\end{equation}
where, ${{\delta }_{1(2)}}={{\omega }_{c}}-{{\omega }_{1(2)}}$  is the frequency detuning between cavity and emitter 1 (emitter 2). The characteristic polynomial in Eq.~(\ref{Eq:8}) and Eq.~(\ref{Eq:9}) is reduced to:

\begin{equation}
{{\left( \varpi -{{\omega }_{c}} \right)}^{3}}+B{{\left( \varpi -{{\omega }_{c}} \right)}^{2}}+C\left( \varpi -{{\omega }_{c}} \right)+D=0.
\label{Eq:11}
\end{equation}
Here, the coefficients of polynomials B, C, D are denoted:

\begin{equation}
\label{Eq:12}
\begin{aligned}
&B={{\delta }_{1}}+{{\delta }_{2}},\\
&C={{\delta }_{1}}{{\delta }_{2}}-{{\kappa }_{1}}{{\kappa }_{2}}+\kappa _{e}^{2}-g_{1}^{2}-g_{2}^{2},\\
&D={{\kappa }_{e}}({{\delta }_{1}}{{\kappa }_{2}}+{{\kappa }_{1}}{{\delta }_{2}})-g_{2}^{2}{{\delta }_{1}}-g_{1}^{2}{{\delta }_{2}}.
\end{aligned}
\end{equation}
Clearly, the pseudo-Hermiticity in the sense of ensure the balanced of gain and loss in the whole system. For simplicity, we take the parameter ratios to replace ${{\kappa }_{2}}$ and ${{g}_{2}}$:
\begin{equation}
{{\kappa }_{1}}=p{{\kappa }_{2}}, {{g}_{1}}=q{{g}_{2}}.
\label{Eq:13}
\end{equation}
We substitute Eq.~(\ref{Eq:13}) into Eq.~(\ref{Eq:10}), these are given by:

\begin{equation}
\label{Eq:14}
\begin{aligned}
&{{\kappa }_{e}}=\left( p+1 \right){{\kappa }_{2}},\\
&{{\delta }_{2}}=-p{{\delta }_{1}},\\
&\delta _{1}^{2}=\frac{{{q}^{2}}+p}{p(p+1)}g_{2}^{2}-\kappa _{2}^{2}.
\end{aligned}
\end{equation}
In addition, we substitute Eq.~(\ref{Eq:14}) into Eq.~(\ref{Eq:12}), we can give:

\begin{equation}
\label{Eq:15}
\begin{aligned}
&B=\left( 1-p \right){{\delta }_{1}},\\
&C=-p\delta _{1}^{2}+({{p}^{2}}+p+1)\kappa _{2}^{2}-({{q}^{2}}+1)g_{2}^{2},\\
&D=(p+1)(1-{{p}^{2}}){{\delta }_{1}}\kappa _{2}^{2}-(1-p{{q}^{2}})g_{2}^{2}{{\delta }_{1}}.
\end{aligned}
\end{equation}
From the last equation in Eq.~(\ref{Eq:14}), it is obvious that it should be satisfied $\delta _{1}^{2}\ge 0$.  To get the minimum value of ${{g}_{2}}$, we may be assumed that $\delta _{1}^{2}=0$, then we can obtain:
\begin{equation}
\label{Eq:16}
{{\left[ {{g}_{2}} \right]}_{\min }}={{\left[ \frac{p(p+1)}{{{q}^{2}}+p} \right]}^{1/2}}{{\kappa }_{2}}.
\end{equation}
In fact, we should satisfy the condition of ${{g}_{2}}\ge {{\left[ {{g}_{2}} \right]}_{\min }}$ in the parameter space.

\section{EP$_{3}$ In the Nanofiber cavity QED}
\label{Sec:3}
In this section, we study the EP$_{3}$ in the symmetric case (${{\kappa }_{1}}={{\kappa }_{2}}$) and asymmetric case (${{\kappa }_{1}}\ne {{\kappa }_{2}}$) by solving the Eq.~(\ref{Eq:11}) under the pseudo-Hermiticity conditions.  First of all, we will find the location where the EP$_{3}$ appears. According to Shengjin formula and the property of EP$_{3}$, the univariate cubic equation has three identical solutions at EP$_{3}$. For simplicity, we assume that $x=\varpi -{{\omega }_{c}}$, therefore the Eq.~(\ref{Eq:9}) can be rewritten as:
\begin{equation}
\label{Eq:17}
{{x}^{3}}+B{{x}^{2}}+Cx+D=0,
\end{equation}
Using the multiple root discriminant, let $a={{B}^{2}}-3C,b=BC-9D$ and $c={{C}^{2}}-3BD$, when $a=b=0$, the solutions ${{x}_{1}}={{x}_{2}}={{x}_{3}}=-\frac{B}{3}=-\frac{C}{B}=-\frac{3D}{C}$.

\subsection{The symmetric case of ${{\kappa }_{1}}={{\kappa }_{2}}$}
\label{Sec:3s}

We consider the ideal case of two identical emitters, which both of them have the same decay rates ${{\kappa }_{1}}={{\kappa }_{2}}$ (i.e., $p=1$) and the same coupling strengths ${{g}_{1}}={{g}_{2}}$ (i.e., $q=1$), the Eq.~(\ref{Eq:14}) and Eq.~(\ref{Eq:15}) can be simplified to:
\begin{equation}
\label{Eq:18}
{{\kappa }_{c}}=2{{\kappa }_{2}};{{\delta }_{2}}=-{{\delta }_{1}};\delta _{1}^{2}=g_{2}^{2}-\kappa _{2}^{2},
\end{equation}
\begin{equation}
\label{Eq:19}
B=0;C=-\delta _{1}^{2}+3\kappa _{2}^{2}-2g_{2}^{2};D=0.
\end{equation}
Then, we can give
\begin{equation}
\label{Eq:20}
a={{B}^{2}}-3C=0\Rightarrow C=4\kappa _{2}^{2}-3g_{2}^{2}=0.
\end{equation}
Therefore, according to Eq.~(\ref{Eq:16}) and Eq.~(\ref{Eq:18}) to Eq.~(\ref{Eq:20}), we can obtain the critical parameters as:
\begin{equation}
\label{Eq:21}
{{\left[ {{g}_{2}} \right]}_{\min }}={{\kappa }_{2}},{{g}_{\text{EP3}}}=\frac{2}{\sqrt{3}}{{\kappa }_{2}},{{\delta }_{\text{EP3}}}=\frac{1}{\sqrt{3}}{{\kappa }_{2}}.
\end{equation}
In addition, the effective pseudo-Hermitian Hamiltonian can be rewritten as:
\begin{equation}
\label{Eq:22}
{{\left( \varpi -{{\omega }_{c}} \right)}^{3}}+\left( 4\kappa _{2}^{2}-3g_{2}^{2} \right)\left( \varpi -{{\omega }_{c}} \right)=0.
\end{equation}
We solve the Eq.~(\ref{Eq:22}) analytically and give the corresponding three eigenvalues:

\begin{equation}
\label{Eq:23}
\begin{aligned}
&{{\varpi }_{1}}={{\omega }_{c}},\\
&{{\varpi }_{2,3}}={{\omega }_{c}}\pm \sqrt{3g_{2}^{2}-4\kappa _{2}^{2}}.
\end{aligned}
\end{equation}
It is worth noting that all our analysis is in the region of $g>{{\left[ {{g}_{2}} \right]}_{\min }}$. Clearly, all of the eigenvalues in Eq.~(\ref{Eq:23}) satisfy this condition. When $g={{g}_{\text{EP3}}}$, we can get ${{\varpi }_{1}}={{\varpi }_{2,3}}={{\omega }_{c}}$ (i.e., EP$_{3}$), which means that three eigenvalues are real. This EP$_{3}$ can be observed in practice because $g>{{\left[ {{g}_{2}} \right]}_{\min }}$.

\subsection{The asymmetric case of ${{\kappa }_{1}}\ne {{\kappa }_{2}}$}
\label{Sec:3y}

In general, it is more difficult that two emitters have the same coupling strengths and decay rates. Therefore, we consider the EP$_{3}$ in the asymmetric case (i.e., ${{\kappa }_{1}}\ne {{\kappa }_{2}}$ and ${{g}_{1}}\ne {{g}_{2}}$). Similarly, assuming that the pseudo-Hermitian system has an EP$_{3}$ at $\varpi ={{\varpi }_{\text{EP3}}}$, it satisfy the following expression as
\begin{equation}
\label{Eq:24}
a={{B}^{2}}-3C=\left( {{p}^{2}}+p+1 \right)\left( \delta _{1}^{2}-3\kappa _{2}^{2} \right)+3({{q}^{2}}+1)g_{2}^{2}=0.
\end{equation}
Combine the Eq.~(\ref{Eq:14}) and Eq.~(\ref{Eq:24}), so the critical parameters are:

\begin{equation}
\label{Eq:25}
\begin{aligned}
&{{\delta }_{\text{EP3}}}={{\left[ \frac{{{q}^{2}}+p}{p(p+1)}g_{\text{EP3}}^{2}-\kappa _{2}^{2} \right]}^{1/2}},\\
&{{g}_{\text{EP3}}}={{\left[ \frac{p+{{q}^{2}}}{p(p+1)}+\frac{3({{q}^{2}}+1)}{{{p}^{2}}+p+1} \right]}^{-1/2}}2{{\kappa }_{2}}.
\end{aligned}
\end{equation}

We give the relation between ${{\left[ {{g}_{2}} \right]}_{\min }}$ and ${{g}_{\text{EP3}}}$ according to the expressions in Eq.~(\ref{Eq:16}) and Eq.~(\ref{Eq:25}):
\begin{equation}
{{g}_{\text{EP3}}}=2{{\left[ {{g}_{2}} \right]}_{\min }}{{\left[ 1+\frac{3p(p+1)\left( {{q}^{2}}+1 \right)}{\left( {{p}^{2}}+p+1 \right)\left( {{q}^{2}}+p \right)} \right]}^{-1/2}}.
\label{Eq:26}
\end{equation}
If we want to give rise to the EP$_{3}$ in the experiments, it is obvious that it must satisfies the condition of ${{g}_{\text{EP3}}}\ge {{\left[ {{g}_{2}} \right]}_{\min }}$. So the parameters $p$ and $q$ should satisfy the following relation:
\begin{equation}
\zeta =\frac{p(p+1)\left( {{q}^{2}}+1 \right)}{\left( {{p}^{2}}+p+1 \right)\left( {{q}^{2}}+p \right)}\le 1.
\label{Eq:27}
\end{equation}

\begin{figure}[!tbp]
\includegraphics[angle=0,width=1\linewidth]{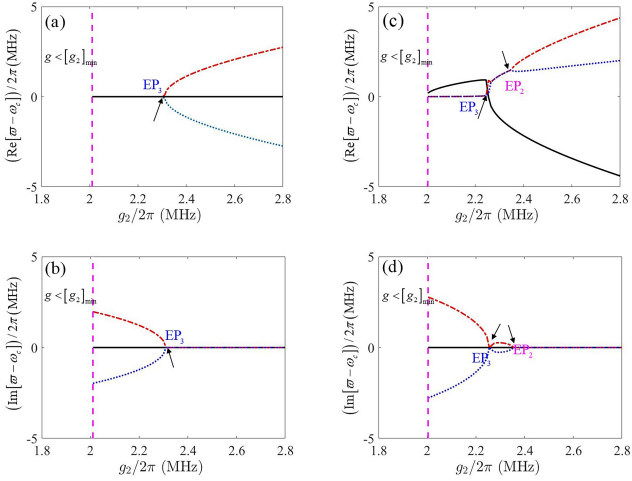}
\caption{(Color online)
 The eigenvalues of the effective Hamiltonian ${{H}_{\text{eff}}}$ in Eq.~(\ref{Eq:6}) as the function of the coupling strength ${{g}_{2}}$ between the cavity mode and the second emitter. The pink dashed lines denote the location of ${{\left[ {{g}_{2}} \right]}_{\min }}$. If ${{g}_{2}}<{{\left[ {{g}_{2}} \right]}_{\min }}$, it will not satisfy the condition of pseudo-Hermiticity. The red dotted lines, blue dotted lines and black solid lines denote the three eigenvalues of ${{H}_{\text{eff}}}$ In panels (a) and (b), we consider both of the emitters have the same coupling strengths and the same decay rates (i.e., $p=1$ and $q=1$), where $\alpha /2\pi =\beta /2\pi =2.25$ MHz and ${{\kappa }_{\operatorname{int}}}/2\pi =0.5$ MHz. In addition, for panels (c) and (d), we discuss the asymmetric case of $p=2$ and $q=2.01$, where $\alpha /2\pi =7$, $\beta /2\pi =1.5$ and ${{\kappa }_{\operatorname{int}}}/2\pi =2.5$. The other parameters of our system are chosen by ${{\kappa }_{2}}/2\pi =2$ MHz.
}
\label{Fig:2}
\end{figure}

In Fig.~\ref{Fig:2}, we plot the eigenvalues of the effective Hamiltonian ${{H}_{\text{eff}}}$ in Eq.~(\ref{Eq:8}) as the function of coupling strength ${{g}_{2}}$ for both the symmetric and asymmetric cases. The left regimes of the pink dashed lines represent ${{g}_{2}}<{{\left[ {{g}_{2}} \right]}_{\min }}$, which have no pseudo-Hermiticity for the system. In Fig.~\ref{Fig:2}(a) and Fig.~\ref{Fig:2}(b), we give the real and imaginary parts of the eigenvalues in Eq.~(\ref{Eq:22}) versus ${{g}_{2}}$ when we choose ${{g}_{1}}={{g}_{2}}$ (i.e., $p=q=1$).
The EP$_{3}$ will be observed at ${{g}_{2}}/2\pi ={{g}_{\text{EP3}}}/2\pi =2.3$ MHz, which result is the consistent with as our calculation based on Eq.~(\ref{Eq:19}). When ${{g}_{2}}>{{g}_{\text{EP3}}}$, all of the three eigenvalues are same and only have the real parts as the black solid lines show. As for when ${{\left[ {{g}_{2}} \right]}_{\min }}\le {{g}_{2}}<{{g}_{\text{EP3}}}$, it is clearly that the eigenvalues have one real and a complex-conjugate pair as the red lines and dotted red lines show. At the critical point (${{g}_{2}}={{g}_{\text{EP3}}}$), the three eigenvalues coalesce.

In addition, from Fig.~\ref{Fig:2}(c) and Fig.~\ref{Fig:2}(d), we discuss the asymmetric condition for ${{g}_{1}}\ne {{g}_{2}}$ (i.e., $p\ne q\ne 1$). Similarly, we plot the corresponding real and imaginary parts of the eigenvalues as the function of coupling strength ${{g}_{2}}$. Choosing the parameters as $p=2$ and $q=2.01$, we prove it satisfy ${{g}_{\text{EP3}}}\ge {{\left[ {{g}_{2}} \right]}_{\min }}$ via $\zeta =0.715< 1$ according to Eq.~(\ref{Eq:25}). We find that when ${{g}_{2}}>{{g}_{\text{EP2}}}$, three eigenvalues only have the real parts. At the EP$_{2}$ as ${{g}_{2}}/2\pi ={{g}_{\text{EP2}}}/2\pi =2.356$,  the complex-conjugate pair are coalescent. When ${{g}_{\text{EP3}}}<{{g}_{2}}<{{g}_{\text{EP2}}}$ and ${{\left[ {{g}_{2}} \right]}_{\min }}\le {{g}_{2}}<{{g}_{\text{EP3}}}$, we find that the eigenvalues consist of a real part and a complex-conjugate pair. Note that near the EP$_{3}$ (${{g}_{\text{EP3}}}/2\pi =2.255$ MHz), the quantum phase transition (i.e., the degeneracy of eigenvalues) will be appearing.

\begin{figure}[!tbp]
\includegraphics[angle=0,width=1\linewidth]{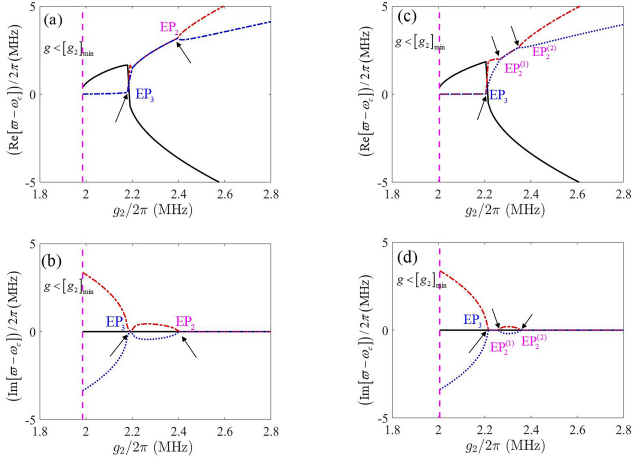}
\caption{(Color online)
 The eigenvalues of the effective Hamiltonian ${{H}_{\text{eff}}}$ when the ratio of two coupling strengths is $p=3$, In panels (a) and (b), we choose the ratio of two dispassion rates $q=3.05$ while we choose $q=3.01$ in panels (c) and (d). Other parameters are the same as those in Fig.~\ref{Fig:2}.
}
\label{Fig:3}
\end{figure}

Furthermore, in order to explore the EP$_{3}$ and EP$_{2}$ induced by the ratio of two coupling strengths $q$, we choose different parameters $q$ with the same parameter $p$ in Fig.~\ref{Fig:3}.  From Fig.~\ref{Fig:3}(a) and Fig.~\ref{Fig:3}(b), by setting the parameters as $p=3$ and $q=3.05$, we plot the real and imaginary parts of the eigenvalues as the function of coupling strength ${{g}_{2}}$, respectively. In these parameters spaces, the parameter $\zeta =0.773< 1$ indicates that ${{g}_{\text{EP3}}}\ge {{\left[ {{g}_{2}} \right]}_{\min }}$ can be realized. These results are similar to the case of $p=2$ [see Fig.~\ref{Fig:3}(a) and Fig.~\ref{Fig:3}(b)], the EP$_{3}$ may be observed at ${{g}_{2}}/2\pi =2.19$ MHz and the EP$_{2}$ will be found at ${{g}_{2}}/2\pi =2.406$ MHz.

Moreover, we choose the parameters $p=3$ and $q=3.01$ in another case, plotting the real and imaginary parts of eigenvalues in Fig.~\ref{Fig:3}(c) and Fig.~\ref{Fig:3}(d), respectively. The parameter $\zeta =0.77< 1$ ensure that ${{g}_{\text{EP3}}}\ge {{\left[ {{g}_{2}} \right]}_{\min }}$ can be realized. Especially, we note that there are three critical coupling strengths: one EP$_{3}$ at ${{g}_{\text{EP3}}}/2\pi =2.216$ MHz and two EP$_{2}$s at ${{g}_{\text{EP}_{2}^{(1)}}}/2\pi =2.256$ MHz and ${{g}_{\text{EP}_{2}^{(2)}}}/2\pi =2.357$ MHz, respectively. These eigenvalues have different characteristics in four regions: when ${{\left[ {{g}_{2}} \right]}_{\min }}\le {{g}_{2}}<{{g}_{\text{EP3}}}$, the eigenvalues have one real and a complex-conjugate pair, and for ${{g}_{\text{EP3}}}<{{g}_{2}}<{{g}_{\text{EP}_{2}^{(1)}}}$ and ${{g}_{2}}>{{g}_{\text{EP}_{2}^{(2)}}}$, the three eigenvalues are same and only have the real parts. When ${{g}_{\text{EP}_{2}^{(1)}}}<{{g}_{2}}<{{g}_{\text{EP}_{2}^{(2)}}}$, the eigenvalues consist of a complex-conjugate pair. In this case, we find that three eigenvalues become coalescent at ${{g}_{2}}={{g}_{\text{EP3}}}$ while the complex-conjugate pair become coalescent at ${{g}_{2}}={{g}_{\text{EP}_{2}^{(1)}}}$ and ${{g}_{2}}={{g}_{\text{EP}_{2}^{(2)}}}$. Most important of all, the transition of eigenvalues degenerate and degenerate broken will happen at these critical points.

In conclusion, compared with the two cases of different parameters $q$ in Fig.~\ref{Fig:3}, we find that via adjusting the coupling strengths between the emitters and the cavity mode, this system will show interesting phenomenon. In a specific parameter space, we can observe a EP$_{3}$ with one or two EP$_{2}$ in this system.

\section{Transmission and absorption spectra In the Nanofiber cavity QED}
\label{Sec:4}

\subsection{CPA conditions}
\label{Sec:4s}

To analysis the CPA conditions, by using the Fourier transformations $a(t)=\frac{1}{\sqrt{2\pi }}\int_{-\infty }^{+\infty }{a(\omega )}{{e}^{-i\omega t}}d\omega $ and $\sigma (t)=\frac{1}{\sqrt{2\pi }}\int_{-\infty }^{+\infty }{\sigma (\omega )}{{e}^{-i\omega t}}d\omega $, we can convert the equations in Eq.~(\ref{Eq:3}) as

\begin{equation}
\label{Eq:28}
\begin{aligned}
&0=i(\omega -{{\omega }_{c}})a-{{\kappa }_{c}}a-i{{g}_{1}}{{\sigma }_{1}}-i{{g}_{2}}{{\sigma }_{2}}+\sqrt{2\alpha }p_{in}^{(1)}+\sqrt{2\beta }p_{in}^{(2)},\\
&0=i(\omega -{{\omega }_{1}}){{\sigma }_{1}}-{{\kappa }_{1}}{{\sigma }_{1}}-i{{g}_{1}}a,\\
&0=i(\omega -{{\omega }_{2}}){{\sigma }_{2}}-{{\kappa }_{2}}{{\sigma }_{2}}-i{{g}_{2}}a.
\end{aligned}
\end{equation}
From the Eq.~(\ref{Eq:26}), We give the intracavity photon number as
\begin{equation}
a(\omega )=-\frac{\sqrt{2\alpha }p_{in}^{(1)}+\sqrt{2\beta }p_{in}^{(2)}}{i(\omega -{{\omega }_{c}})-{{\kappa }_{c}}+\sum{(\omega )}}.
\label{Eq:29}
\end{equation}
Here,
\begin{equation}
\sum{(\omega )}=\frac{g_{1}^{2}}{i(\omega +{{\delta }_{1}}-{{\omega }_{c}})-{{\kappa }_{1}}}+\frac{g_{2}^{2}}{i(\omega +{{\delta }_{2}}-{{\omega }_{c}})-{{\kappa }_{2}}}
\label{Eq:30}
\end{equation}
is the self-energy due to the two emitters.

Using the input-output relation, we can obtain the output fields $p_{out}^{(1)}$ and $p_{out}^{(2)}$ at ports 1 and 2

\begin{equation}
\label{Eq:31}
\begin{aligned}
p_{out}^{(1)}=\sqrt{2\alpha }a(\omega )-p_{in}^{(1)},\\
p_{out}^{(2)}=\sqrt{2\beta }a(\omega )-p_{in}^{(2)}.
\end{aligned}
\end{equation}
According to Eq.~(\ref{Eq:5}), we can obtain the relationship of the input fields $p_{in}^{(1)}$ and $p_{in}^{(2)}$ as
\begin{equation}
\frac{p_{in}^{(2)}}{p_{in}^{(1)}}=\sqrt{\frac{\beta }{\alpha }},
\label{Eq:32}
\end{equation}
which means that the two input laser fields should share the same phase with strength ratio as $\sqrt{\frac{\beta }{\alpha }}$.
Solving Eq.~(\ref{Eq:31}) with $p_{out}^{(i)}=0$, the other two equations in the parameters space and the frequency of the two input fields are

\begin{equation}
\label{Eq:33}
\begin{aligned}
&{{\kappa }_{c}}=\sum\limits_{i=1}^{2}{\frac{g_{i}^{2}}{{{\left( {{\omega }_{i}}-{{\omega }_{\text{CPA}}} \right)}^{2}}+\kappa _{i}^{2}}}{{\kappa }_{i}},\\
&{{\omega }_{c}}-{{\omega }_{\text{CPA}}}=\sum\limits_{i=1}^{2}{\frac{g_{i}^{2}}{{{\left( {{\omega }_{i}}-{{\omega }_{\text{CPA}}} \right)}^{2}}+\kappa _{i}^{2}}}\left( {{\omega }_{i}}-{{\omega }_{\text{CPA}}} \right),
\end{aligned}
\end{equation}
where ${{\omega }_{\text{CPA}}}$ is the frequency of the input fields in the cases of CPA.

\subsection{The output spectrum}
\label{Sec:4x}

In this section, we plot the total output spectrum of the cavity for this nanofiber cavity QED and show that the EP$_{3}$ and EP$_{2}$ can also be observed in the output spectrum. According to the relation of the two input fields in Eq.~(\ref{Eq:32}), the expressions of the two output fields can be written as

\begin{equation}
\label{Eq:34}
\begin{aligned}
&p_{out}^{(1)}={{S}_{1}}(\omega )p_{in}^{(1)},\\
&p_{out}^{(2)}={{S}_{2}}(\omega )p_{in}^{(2)},
\end{aligned}
\end{equation}
Thus, we define the total output spectrum which is measured from the right port of the system.
\begin{equation}
\label{Eq:35}
{{S}_{1}}(\omega )={{S}_{2}}(\omega )=-1-\frac{2\alpha +2\beta }{i(\omega -{{\omega }_{c}})-{{\kappa }_{c}}+\sum{(\omega )}},
\end{equation}
And we define the total output spectrum as ${{\left| S(\omega ) \right|}^{2}}={{\left| {{S}_{1}}(\omega ) \right|}^{2}}+{{\left| {{S}_{2}}(\omega ) \right|}^{2}}$.  It is easy to check that when $\omega ={{\omega }_{{CPA}}}$ in Eq.~(\ref{Eq:33}), the total output spectrum ${{\left| S(\omega ) \right|}^{2}}=0$.

\begin{figure}[!tbp]
\includegraphics[angle=0,width=1\linewidth]{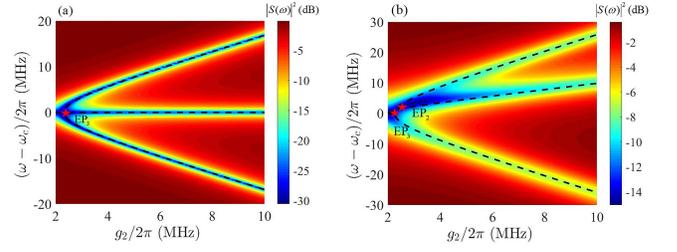}
\caption{(Color online)
The total output spectrum ${{\left| S(\omega ) \right|}^{2}}$ as the function of the coupling strength ${{g}_{2}}$ between the cavity mode and the second emitter and the frequency detuning $\omega -{{\omega }_{c}}$ between the input field and cavity mode. In panel (a), we discuss the symmetric case of $p=1$ and $q=1$. In panel (b), we study the asymmetric case of $p=2$ and $q=2.01$. The three eigenvalues are indicated by using the black dashed lines. The exceptional points (EP$_{3}$ and EP$_{2}$) in Fig.~\ref{Fig:2} can also be found in the two output spectrums, as the red stars show. Other parameters are the same as those in Fig.~\ref{Fig:2}.
}
\label{Fig:4}
\end{figure}

\subsection{The transmission and absorption spectrum}
\label{Sec:4y}
In order to study the interesting phenomenon induced by ${{g}_{\text{EP3}}}$, we show the transmission spectra and absorption spectra of the system in Fig.~\ref{Fig:4}.
By using the steady-steady intracavity field in Eq.~(\ref{Eq:29}) and combine the coherent input field via the input-output formalism, the transmission spectra can be calculated as
\begin{equation}
\label{Eq:36}
T=\frac{{{\left| p_{in}^{(2)}+\sqrt{2\beta }a(\omega ) \right|}^{2}}}{{{\left| p_{in}^{(1)} \right|}^{2}}+{{\left| p_{in}^{(2)} \right|}^{2}}},
\end{equation}
In Fig.~\ref{Fig:4}(a), we plot the transmission spectrum versus the frequency detuning $\omega -{{\omega }_{c}}$ in the simplest condition of ${{g}_{1}}={{g}_{2}}$ and ${{\kappa }_{1}}={{\kappa }_{2}}$. we consider five cases with different coupling strengths ${{g}_{2}}$ and explain this phenomenon in three conditions. Correspondingly, the transmission spectra have been given in Fig.~\ref{Fig:4}(b). First, we set that ${{\left[ {{g}_{2}} \right]}_{\min }}<{{g}_{2}}<{{g}_{\text{EP3}}}$ in the cases of (i) and (ii), the only single peaks at central positions (see the red and green lines) can be observed. Second, at the critical point (${{g}_{2}}={{g}_{\text{EP3}}}$) as shown in the case of (iii), the result is same to those in the first two cases (see the blue line). Third, in the case of (iv) and (v), we set that ${{g}_{2}}>{{g}_{\text{EP3}}}$, it is clearly that three peaks can be observed. And the new symmetric peaks arise (as the black arrow shows) when we send a resonant ($\omega ={{\omega }_{c}}$) input field (see the pink and black lines). We note that with increase of the coupling strength ${{g}_{2}}$, the splitting of this central peak become more pronounced. And a stronger emitter-cavity coupling strength indicates a large reduction in transmission.

To explain this phenomenon, it is convenient to show the physical mechanism of the system by using the collective states. Due to the whole system is in the weak driving limit, so we write these states in one-photon space \cite{PhysRevA.95.063842}: $\left| gg,1 \right\rangle $ and $\left| \pm ,0 \right\rangle $, where $\left| \pm  \right\rangle =\frac{1}{\sqrt{2}}\left( \left| eg \right\rangle \pm \left| ge \right\rangle  \right)$.
Using the Hamiltonian in Eq.~(\ref{Eq:1}), we can obtain a set of eigenvalues ${{\lambda }_{0}}={{\omega }_{c}}\hbar $ and ${{\lambda }_{\pm }}={{\omega }_{c}}\hbar \pm \sqrt{2}g\hbar $.
Clearly, when ${{\left[ {{g}_{2}} \right]}_{\min }}<{{g}_{2}}\le {{g}_{\text{EP3}}}$, this system is in the dressed states $\left| gg,1 \right\rangle $ with the eigenvalues ${{\lambda }_{0}}$, corresponding to the single central peaks. And we observed broadening of the spectra for the cases of (i)-(iii) with the increase of coupling strength ${{g}_{2}}$. When ${{g}_{2}}>{{g}_{\text{EP3}}}$, the system is in the states $\left| gg,1 \right\rangle $ and $\left| \pm ,0 \right\rangle $ with the eigenvalues ${{\lambda }_{0}}$ and ${{\lambda }_{\pm }}$. Therefore, we can observe the additional symmetric peaks in the two sides.

\begin{figure}[!tbp]
\includegraphics[angle=0,width=1\linewidth]{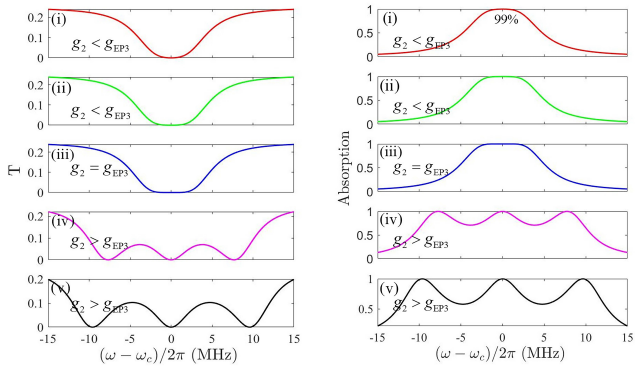}
\caption{(Color online)
The total transmission spectra and the absorption spectra as functions of frequency detuning $\omega -{{\omega }_{c}}$ are shown in panel (a) and panel (b), respectively. The coupling strength ${{g}_{2}}$ between the cavity mode and the second emitter are setting $2\pi \times \left( 2,2.15,4/\sqrt{3},5,6 \right)$ MHz for (i)-(v), respectively. The black arrows show the new peaks at the central position. The parameters are same to those in Fig.~\ref{Fig:2}.
}
\label{Fig:5}
\end{figure}

\section{EXPERIMENTAL IMPLEMENTATION}
\label{Sec:5}

Now, we will discuss the feasibility of this present configuration to observe the higher-order exceptional points. The cavity length can reach the order of meters by measuring the free spectral range \cite{PhysRevLett.115.093603}. In addition, the cavity-field decay rates through the FBG$i$ mirrors are given by
$\alpha =\frac{c}{4l}(1-{{R}_{1}})$ [$\beta =\frac{c}{4l}(1-{{R}_{2}})$]. And total cavity field decay rate is given by
${{\kappa }_{c}}=\frac{c}{4l}\left[ (1-{{R}_{1}})+(1-{{R}_{2}})+2\eta  \right]$, where $l$ is the length of the nanofiber cavity, ${{R}_{i}}$ is reflectance of the FBG$i$ mirrors and $\eta $ the intrinsic cavity loss \cite{PhysRevApplied.13.064010}. Thus, our system exhibits good tunability by adjusting length of the nanofiber cavity and the FBG$i$ mirrors reflectance.

The emitters can be loaded into the nanofiber trap from a standard six-beam optical molasses \cite{PhysRevLett.110.243603}. From the above analysis, we noticed that it is important to tune the ratio parameter $q$ of the coupling strengths. We can place the two identical atoms at the same distance from the nanofiber surface to achieve ${{g}_{1}}={{g}_{2}}$ (i.e., $q=1$), which can realize experimentally. As for realizing different coupling strengths ${{g}_{1}}\ne {{g}_{2}}$ (i.e., $q\ne 1$),
we can choose two different atoms into the system. In addition, the positions of the emitters ${{z}_{i}}$ are tunable so that the position-dependent emitter-cavity coupling strength is given by $g_{i}^{'}={{g}_{i}}\cos (2\pi {{z}_{i}}/{{\lambda }_{c}})$, where ${{\lambda }_{c}}$ being the wavelength of the nanofiber cavity mode.

CPA has been achieved in an optical system without PT symmetric \cite{RN164}. In our work, we realize CPA with pseudo-Hermiticity in an all-fiber emitter-cavity QED system.
The CPA conditions can be perfectly satisfied by adjusting the parameters of the system, where the absorption rate of the cavity for the input fields can reach 99\%
as shown in Fig.~\ref{Fig:5}(b). Compared with other work, the CPA condition with pseudo-Hermiticity and the observation of EP$_{3}$ may be realizable in the experiment via using our method.

\section{CONCLUSION}
\label{Sec:6}

In summary, we present a nanofiber cavity-emitter QED system and theoretically study the pseudo-Hermiticity and EP$_{3}$ in this system. In the appropriate parameter spaces of the pseudo-Hermiticity, the effective Hamiltonian of the system has three eigenvalues, which including the cases of three real eigenvalues and one real part with a complex-conjugate pair. By tuning the two emitter-cavity coupling strengths, the three eigenvalues are coalescent at the EP$_{3}$. In addition, the system can be discovered three critical points (one EP$_{3}$ and two EP$_{2}$s) via tuning the ratio of the coupling strengths. Furthermore, we plot the output spectrum and transmission spectra to demonstrate the existence of the EP$_{3}$ and reveal the pseudo-Hermiticity of the system. We analysis the experimental feasibility realize the pseudo-Hermiticity and EP$_{3}$ in an all-fiber emitter-cavity QED system under reasonable parameter interval. These characteristics may provide a new platform to achieve high-order exceptional points in a large-scale all-fiber quantum network.

\begin{acknowledgments}

This work was financially supported by the NationalNatural Science Foundation of China (11804018 and 62075004 )，and Beijing Natural Science Foundation (4212051).
\\
\end{acknowledgments}

\end{CJK}
\end{document}